# Fast IDS Computing System Method and its Memristor Crossbar-based Hardware Implementation

Sajad Haghzad Klidbary, *Student Member, IEEE*, Saeed Bagheri Shouraki, and Iman Esmaili Pain Afrakoti

*Abstract*—**Active Learning Method (ALM) is one of the powerful tools in soft computing that is inspired by human brain capabilities in processing complicated information. ALM, which is in essence an adaptive fuzzy learning method, models a Multi-Input Single-Output (MISO) system with several Single-Input Single-Output (SISO) subsystems. Ink Drop Spread (IDS) operator, which is the main processing engine of this method, extracts useful features from the data without complicated computations and provides stability and convergence as well. Despite great performance of ALM in applications such as classification, clustering, and modelling, an efficient hardware implementation has remained a challenging problem. Large amount of memory required to store the information of IDS planes as well as the high computational cost of the IDS computing system are two main barriers to ALM becoming more popular. In this paper, a novel learning method is proposed based on the idea of IDS, but with a novel approach that eliminates the computational cost of IDS operator. Unlike traditional approaches, our proposed method finds functions to describe the IDS plane that eliminates the need for large amount of memory to a great extent. *Narrow Path* and *Spread*, which are two main features used in the inference engine of ALM, are then extracted from IDS planes with minimum amount of memory usage and power consumption. Our proposed algorithm is fully compatible with memristor-crossbar implementation that leads to a significant decrease in the number of required memristors (from $O(n^2)$ to $O(3n)$). Simpler algorithm and higher speed make our algorithm suitable for applications where real-time process, low-cost and small implementation are paramount. Applications in clustering and function approximation are provided, which reveals the effective performance of our proposed algorithm.**

## I. INTRODUCTION

Nowadays, various scientific areas, including artificial neural networks and fuzzy logic, are trying to discover and simulate the way in which human brain processes information.

On one hand, artificial neural networks investigates the neural network of living organisms; fuzzy logic, on the other hand, studies the functional properties of the human brain. These tools have been widely reported with numerous successful applications in function approximation, time series forecasting, and data mining; however, fuzzy logic has special status[1]. The term fuzzy logic was first introduced by Lotfi A. Zade in 1965[2]. In system modelling, classic methods describe the system by complicated mathematical equations; whereas fuzzy logic employs a different approach and describes the system with a set of linguistic "IF-THEN" expressions. With this approach, fuzzy logic not only relieves the accurate computation burden, but also performs better in the presence of uncertainty[3].

In 1997, Shouraki *et al.* introduced Active Learning Method (ALM) which is one of the powerful algorithms in soft computing[4], [5], [6]. ALM is a modelling and control algorithm which is based on learning capabilities and expertise of the human brain and tends to find a logical cause-and-effect relationship between events. It was first employed to control an inverted pendulum [7]. Despite its counterparts in modelling (fuzzy algorithms like Sugeno-Yasukawa[8] and Takagi-Sugeno[9] and neuro-fuzzy algorithms like ANFIS[10]), ALM enjoys a simpler structure and faster training phase [5]. Avoiding time consuming, iterative and complicated computations along with fast learning, stability and noise resistance [11] are some advantages of this algorithm. ALM has been successfully employed in numerous applications in control[7],[12],[13], robotics[14],[15], modelling [11], soft computing and artificial intelligence[16], image processing [17] and real-time processing[15].

ALM is inspired by some hypotheses which claim that the human brain interprets information in pattern-like images rather than numerical quantities and tends to break down complex problems into some simpler subproblems. The main idea of ALM is to approximate a Multi-Input Multi-Output (MISO) system with several Single-Input Single-Output (SISO) subsystems. Each of these subsystems, which is modelled by an Ink Drop Spread (IDS) plane, represents the relationship of the output with respect to one of the inputs of the main system. Two main features called *Narrow Path* and *Spread* are extracted from each IDS plane. These two features from all IDS planes are then fed to the inference engine in order to approximate the output. The number of required IDS planes is determined by the number of inputs and the number of fuzzy partitions on each as well as the complexity of hardware implementation. Therefore the bottleneck is an efficient hardware implementation of IDS operator. Different implementations have been proposed, Murakami in [18] proposed a digital implementation which requires large amount of memory and a high-level controller to manage data transfer. Pipeline implementation by Firouzi [19] suffers from digital computing problems such as overflow and finite precision. Tarkhan's analog implementation, however, suffers from high power consumption and high complexity of each IDS plane [20].

Memristor is a nonlinear passive two-terminal electrical element whose resistance is controlled by voltage and can act as a memory resistor. Because of the compatibility of IDS planes with memristor-crossbar structure, the implementation of IDS planes on this structure is efficient and of low cost. For this reason, Merrikh-bayat proposed the first memristor crossbar hardware implementation of IDS planes [21] and then, a more efficient implementation was proposed by Esmaili [22]; however, both implementations suffer from large hardware problem. These hardware implementations suffer from large amount of memory required to store the information of IDS planes and consequently higher power consumption. For instance, for an IDS plane, $m = R_{sn_x} \times R_{sn_y}$ memory cells (memristors) are required, where $R_{sn_x} \times R_{sn_y}$ is the resolution of the IDS plane. This large number of memristors are enough to implement the information of only one IDS plane, and as the number of inputs or their fuzzy partitioning increases, the number of IDS planes increases inevitably. In addition, the hardware required to implement IDS operator and inference engine also add to the complexity of the overall hardware implementation.

In our proposed algorithm, a novel approach to describe IDS planes is employed, which eliminates the computational complexity of IDS operator and requires only $m = 3 \times R_{sn_x}$ memory cells (memristors) for each IDS plane. This results in significant reduction in the hardware complexity of the memory unit from $O(n^2)$ to $O(3n)$, where $n$ is the resolution of an IDS plane on each of its axis. In our proposed algorithm three rows of memory cells with resolution of $R_{sn_x}$ memristors are assumed, two of which store the upper-bound and the lower-bound of the input-output relationship in an IDS plane. The difference between these two rows (or vectors) is inversely proportional to the degree of belief by which the associated input value is observed, and the last row keeps the values of the output. In the learning phase, when a new training sample is observed, each element of these three vectors will be updated by applying a coefficient of their distance to the new observed sample.

Following are justifications of our proposed algorithm:

1) The large number of required memory cells to store the information of IDS planes as well as partial usage of the whole memory in traditional approaches reveals the need for a novel approach, particularly in the way IDS planes are described.

2) The large hardware required to implement IDS planes and IDS operator should be optimized to make it suitable for real-time applications.

3) Hardware should be reduced in size and complexity to optimize the power consumption.

4) In traditional approaches, there was a need for a diode with low inverse current to be serried with memristor and to reduce the feedback effect. The proposed approach eliminates this need.

5) In the traditional approaches, the memristive implementation was based on approximated mathematical equations; whereas, the proposed hardware is implemented with accurate mathematical equations.

Therefore the proposed algorithm enjoys a significant reduction in the large amount of required memory and in the computational complexity of the IDS operator. Due to the analog mechanism of our proposed algorithm, it out performs digital implementations in terms of learning and test speed. In addition, the proposed hardware is implementable with considerably lower computational complexity.

The remainder of this paper is organized as follows. The main concepts of ALM and IDS operator are reviewed in section 2, followed by a brief description of the forth fundamental electrical element called memristor in section 3. Section 4 illustrates our proposed algorithm and its memristor-crossbar hardware implementation is presented in section 5. The evaluation of our proposed algorithm is presented in section 6. Eventually section 7 concludes our paper and provides suggestions for future work.

## II. ACTIVE LEARNING METHOD (ALM)

Processes and computations in the human brain are believed to be qualitative and imprecise in their essence. This observation has led to a research field called soft computing, which deals with uncertainty and simulates the human brain. Among those, fuzzy systems, which is inspired by human brain capabilities, can be employed as a tool to deal with uncertainty and to provide stability in real-world problems. ALM has adopted a fuzzy approach and has its basis in hypotheses which claim that humans interpret their surrounding environment in an inexact manner and rather than dealing with quantities and numbers, a general characteristic of the environment is learned. Facing a new problem, humans tend to avoid delving into details; instead, a general understanding of the problem is preferred and, if required, minor problems are overcome first. Therefore, in dealing with complicated problems, humans first attempt to find simpler or familiar concepts and then by discovering logical connections among these concepts, obtain an inexact definition and a general understanding of the problem without much effort. ALM is an adaptive fuzzy learning method that obtains a clearer understanding of the original problem by splitting a complicated problem into several simpler ones. ALM approach to splitting a MISO system into several SISO subsystems is illustrated in Fig. 1.

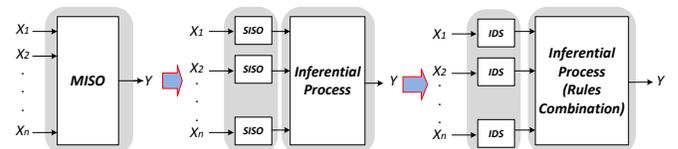

Fig. 1. ALM approach to splitting a MISO system into several SISO subsystems. For each SISO subsystem, an IDS unit is allocated to extract the features. All features are then aggregated in the inference engine to approximate the output.

ALM has two main steps. The first step is updating IDS planes when training samples are observed and the last step in which inference process is done. These steps are discussed in the following subsections.

### A. Ink Drop Spread (IDS) Operator

The heart of ALM is the IDS operator which somehow models fuzzy interpolation. Despite traditional learning algorithms in which system behavior is represented by complicated mathematical equations, ALM tries to simulate human brain functionality by providing a qualitative and behavioral description of the system. In ALM, the imprecise and fuzzy characteristics of the human brain in learning from events is modeled by IDS operator. The function of the IDS operator is inspired by the fact that experiences in the data space are continuous in essence. In other words, the space of learning is not confined to those observed samples; moreover, in the vicinity of an observed sample, the learned features are still valid, though, probably with less certainty as we move away from that observed sample. If the domain of all inputs and output are quantized, then an IDS plane associated with a SISO subsystem $(X_i, Y)$ is a gridded plane that depicts the projected samples of input $x_i$ and output $y$ for a specific domain of variation of other inputs. The effect that the IDS operator has on an observed sample is analogous to instilling an ink drop on the coordinate of that sample. The IDS operator is applied to all observed samples of input $x_i$ and output $y$ on the $X_i - Y$ IDS plane. Each IDS unit is comprised of two main parts. A 2-D plane that captures the $X_i - Y$ relationship and the Feature Extracting Unit that extracts useful information from the formed pattern on the IDS plane as shown in Fig. 2.

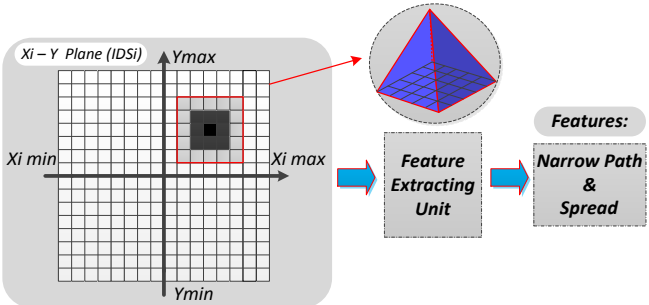

Fig. 2. The structure of an IDS unit. Initially the IDS plane is white (empty) and when a new sample is observed the associated ink drop will spread on the IDS plan, as an ink drop spreads on a sheet of paper. *Narrow Path* and *Spread* are two features that will be extracted from the pattern formed on the IDS plane by feature extraction unit. These features are then fed to the inference engine. In this figure a pyramid membership function is assumed.

For simplicity the number of IDS planes is assumed to be equal to the number of inputs (no partitioning is performed on inputs domain). When a training sample $x_i$ is observed, where $1 < i < N$, and $N$ is the size of the training dataset. The $IDS_j$ plane, which is associated with the relationship between the output and the $j$th input, is updated according to the projection of that sample on this IDS plane. For instance, assume two distinct samples $P_1(x_1^{train}, x_2^{train}, y) = (-5,4,5)$ and

$P_2(x_1^{train}, x_2^{train}, y) = (4, -1,4)$ are observed. In this two-input single-output system, two white (empty) IDS planes $X_1 - Y$ and $X_2 - Y$ are assumed. As shown in Fig. 3 $P_{11}^{IDS1}(x_1^{train}, y) = (-5,5)$ and $P_{21}^{IDS1}(x_1^{train}, y) = (4,4)$ spread on the $IDS_1$ plane. Meanwhile, $P_{12}^{IDS2}(x_2^{train}, y) = (4,5)$ and $P_{22}^{IDS2}(x_2^{train}, y) = (-1,4)$ spread on the $IDS_2$ plane. The membership function for spreading ink drops can be considered Gaussian, Pyramid, Conic or any convex 3-D shape that decreases as the distance to its centers increases. In Fig 3 with pyramid membership function, as ink drops spread, overlapped patterns become increasingly darker. These darker regions increase our belief in those regions of the input space.

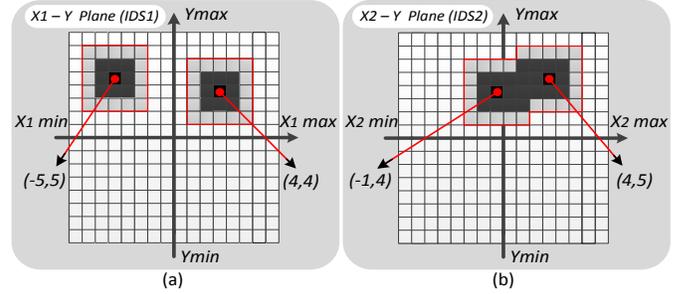

Fig. 3. An example of 2-input single-output system with pyramid membership function and two training samples $P_1(x_1^{train}, x_2^{train}, y) = (-5,4,5)$, $P_2(x_1^{train}, x_2^{train}, y) = (4, -1,4)$. (a) The $IDS_1$ plane associated with $X_1 - Y$ relationship where the projection of two training samples is spread $P_{11}^{IDS1}(x_1^{train}, y) = (-5,5)$, $P_{21}^{IDS1}(x_1^{train}, y) = (4,4)$. (b) The $IDS_2$ plane associated with $X_2 - Y$ relationship where the projection of two training samples is spread $P_{12}^{IDS2}(x_2^{train}, y) = (4,5)$, $P_{22}^{IDS2}(x_2^{train}, y) = (-1,4)$. As it can be seen, the overlapped regions have become increasingly darker. The increased darkness of some grids signifies the correctness and support of other observed samples. The overlapping ink drops somehow models fuzzy interpolation.

Fig. 4 shows a formed pattern on an IDS plane when all training samples have been observed and their associated ink drops have spread. The method to find the values of *Narrow Path* and *Spread* in the position of a candidate point $x_i^*$ are shown on the figure. The *Narrow Path* function for the formed pattern has been shown with red color and the inverse of *Spread* around the *Narrow Path* reveals the degree of certainty regarding the occurrence of that observation. One of the advantages of ALM is that only one epoch is sufficient to learn the training set and no initialization is required. The order of observing training samples does not affect the final result and the algorithm gradually learns through interaction with the system. When a new training sample is observed only local regions, rather than the whole plane, are updated[11]. It is worth mentioning that the radius, an ink drop is permitted to spread, is an important parameter that affects the performance, convergence speed and output error. This user-defined parameter is inversely correlated with the size of the training set and the density of samples. This parameter is defined according to the resolution of the IDS plane as well as the density and distribution of training samples. In the original version of the ALM this parameter is chosen through an iterative trial and error approach which is time consuming. In the figure provided, it can be seen that the IDS plane can be treated as a 2-D memory unit, in which each cell stores the darkness of the associated grid on the IDS plane.

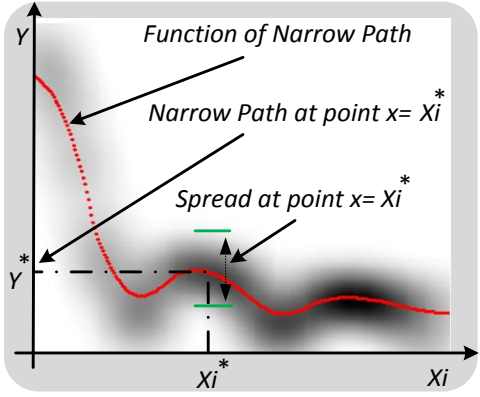

Fig. 4. An example of a typical pattern formed on $IDS_j$ plane when the IDS operator is applied to all training samples on jth plane. The *Narrow Path* function is shown with red color and the method to find the *Narrow Path* and Spread in the position of a candidate point $x_i^*$ is shown on the figure. The inverse of *Spread* around an input value reveals the degree of belief we have regarding the occurrence of that observation.

### B. Inference Engine in ALM

After the IDS operator has been applied to all IDS planes and the *Narrow Path* and *Spread* for all IDS planes have been extracted, these features are then fed to the inference engine of the ALM. The *Narrow Path* function in the $IDS_i$ reveals the relationship between the output and the $ith$ input and the value of *Spread* in this plane, compared to spread in other planes, is an indication of the importance of the $ith$ input $x_i$ in approximating the output. As shown in Fig. 4, the *Spread* in some input domains is wider than that of other domains. Wide *Spread* in some input domains, implies that in those input domains the output is more affected by other inputs rather than $x_i$, because the output has varied a lot, while the $ith$ input held almost the same value. This is the justification for how a MISO can be approximated by several SISO subsystems in ALM.

If the *Spread* is wide for all input domains on an IDS plane, fuzzy partitioning the universe of discourse of other inputs is proposed. In this case, for each partition a separate IDS plane is required and by splitting the input domain more knowledge can be extracted from IDS planes, which leads to lower approximation error. It should be noted that if the datasets is small, excessive partitioning the input domain will result in high approximation error.

Various approaches have been proposed to find the *Narrow Path* including maximum operator and averaging operator. In order to measure the *Spread,* the width of the ink drop spread around each point can be employed. The following is an approach proposed in[11]. It should be mentioned that because of the fuzzy and inexact nature of the IDS operator in ALM, different approaches for extracting these features do not vary that much in final results and the selection among different approaches depends on the hardware implementation and processing speed considerations.

For the dataset $S$ that contains $N$ training samples:

$$S = \{(x_1, y_1), (x_2, y_2), \dots, (x_N, y_N)\} \qquad (1)$$

Where each input is in $D$ dimensional space and we will have:

$$x = (x_1, x_2, \dots, x_D)^T \qquad , \quad x \subseteq R^D \qquad (2)$$

The first step of ALM is quantizing the input and output domains in each IDS plane. For simplification, quantization levels for both input and output can be the same and equal to $R_{sn}$. By this choice the resolution of the IDS plane is $R_{sn_x} \times R_{sn_y}$, where $R_{sn_x} = R_{sn_y}$. Therefore, we have:

$$X_i = \begin{cases} \left\lfloor \dfrac{(x_i - x_{min}^i) * Rsn_x}{x_{max}^i - x_{min}^i} \right\rfloor + 1 & : \ x_i \in (x_{min}^i, x_{max}^i) \\ 1 & : \ x_i \leq x_{min}^i \\ Rsn_x & : \ x_i \geq x_{max}^i \end{cases} \qquad (3)$$

$$Y = \begin{cases} \left\lfloor \dfrac{(y - y_{max}) * Rsn_y}{y_{max} - y_{min}} \right\rfloor + 1 & : \ y \in (y_{min}, y_{max}) \\ 1 & : \ y \leq y_{min} \\ Rsn_y & : \ y \geq y_{max} \end{cases} \qquad (4)$$

Where $1 \leq i \leq D$ and $D$ is the number of IDS planes. The quantized values are as follows:

$$X_i \in \{1, 2, \dots, Rsn_x\} \qquad (5)$$

$$Y \in \{1, 2, \dots, Rsn_y\} \qquad (6)$$

If $p(x, y)$ is a projected point to the $y - x_i$ space and we assume that its darkness is indicated by $d(x, y)$ and its membership function is a Gaussian with maximum of one and appropriate variance, then observing such a sample entails updating the $(x_s, y_s)$ grid on the $IDS_i$ plane as follows:

$$P_{x,y} = \{p(x, y) | x \in X_i, y \in Y\} \qquad (7)$$

$$d(x_s + u, y_s + v) = d(x_s + u, y_s + v) + h(u, v)$$
$$, -R \leq u \ , \ v \leq R \qquad (8)$$

Where $R$ indicated the radius of ink drop spread and $h$ is the shape of ink drop spread. With regard to definitions of *Narrow Path* and *Spread*, these two features are calculated as follows:

$$\psi_{x_i}(x) = \left\{ b \mid \sum_{y=y_{min}}^{b} d(x, y) \approx \sum_{y=b}^{y_{max}} d(x, y) \ , b \in Y \right\} \qquad (9)$$

$$S_{x_i}(x) = \max_{y \in Y} \{y | d(x, y) > T\} - \min_{y \in Y} \{y | d(x, y) > T\} \qquad (10)$$

Where $\psi_{x_i}$ and $S_{x_i}$ are *Narrow Path* and *Spread* on the $IDS_i$ plane respectively. The first equation implies that the *Narrow Path* value on the $IDS_i$ plane for any given quantized input $x$ is $b$, if the sum of the grids darkness values above the $(x, b)$ grid is approximately equal to the sum of the grids darkness values below the $(x, b)$ grid. The second equation implies that the *Spread* value on the $IDS_i$ plane for any given quantized input $x$ is proportional to the effective width of the formed pattern on the column of grids on the coordination of $x$. In this equation, the parameter $T$ indicated the minimum acceptable darkness of a grid on the IDS plane to measure the *Spread*, and is defined by the user. When a new test sample is observed, $x = (x_1, x_2, ..., x_D)$, where $D$ is the dimension of the test sample, the *Narrow Path* and *Spread* values for this sample from all IDS planes are extracted and then summed to approximate the output as follows:

$$y(x) = \sum_{i=1}^{c} \beta_i \psi_{x_i}(x_i) \tag{11}$$

$$\beta_i = \frac{\frac{1}{S_{x_i}(x_i)}}{\sum_{j=1}^{N} \frac{1}{S_{x_j}(x_j)}} \tag{12}$$

Where $c$ is the number of IDS planes (if no partitioning is performed, also the number of inputs). As it can be seen, the inference in ALM is done by weighted sum of *Narrow Paths* where $\beta_i$ is the weight associated with the $i$th *Narrow Path*, which is the normalized value of *Spread* inverse or any descending function (the wider the *Spread* in $IDS_i$ plane the lower we believe in the *Narrow Path* of $i$th plane and vice versa). For those values of $x$ whose $y$ grid on the IDS plane is white, $\psi_{x_i}(x)$ and $S_{x_i}(x)$ values are assumed to be equal to $y_{max}/2$.

Fig. 5 shows the flowchart of ALM algorithm. In the first step of this algorithm for those IDS planes where data samples are sparse, data sampling is performed intelligently. Initially no partitioning on inputs domain is performed. After IDS operator is applied and *Narrow Path* and *Spread* are extracted, the most effective inputs will be identified. The model is then built and if the approximation error is within a user-defined range, the algorithm stops; otherwise, inputs domain partitioning will be performed. Even if by partitioning the inputs domain, samples on IDS planes are still sparse and the algorithm is not successful to decrease the error, the algorithm returns to the first step and samples more data and rebuilds a new model to decrease the error. In the next step, if the *Spread* in an IDS plane is greater than a user-defined threshold, the algorithm intelligently identifies that IDS plane and more data sampling is performed for that particular IDS plane. Various methods for partitioning the input domain have been proposed, one option is to double the partitioning in each step[11].

In Aristotle's logic, in order to enhance knowledge, it is proposed that more details should be extracted, this is in stark contrast with ALM which is inspired by human brain functionality. When faced with a new learning situation, the human brain tends to discard details and learn the overall behavior. In ALM, if desirable and sufficient knowledge is not achieved, then by splitting the original system into some simpler subsystems, more knowledge is aimed to achieved within each SISO subsystem, because each of which only deals with a particular domain of inputs. Eventually, in the inference unit of ALM, the extracted features, *Narrow Path* and *Spread*, from all IDS planes are aggregated to approximate the output. The number of fuzzy rules in inference unit is equal to the number of IDS planes.

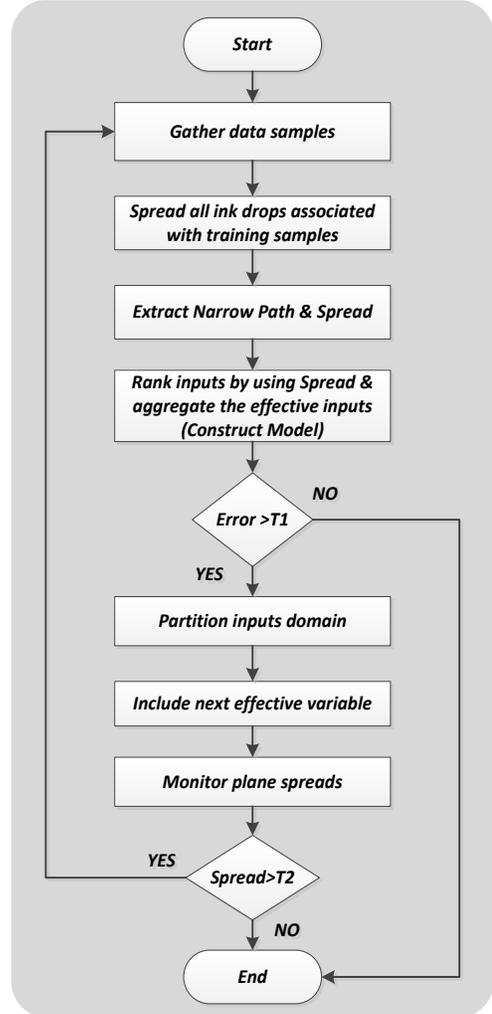

Fig. 5. ALM flowchart. In this algorithm the most effective inputs in approximating the output are intelligently identified and the algorithm tries to build a model with the minimum complexity and the number of inputs. In this flowchart, T1 is the desirable error threshold and T2 is the threshold that defines the minimum acceptable data density on IDS planes.

## III. MEMRISTOR

In addition to the three previously known fundamental electrical elements: resistor, capacitor and inducer, in 1971, Leon Chua mathematically proved and introduced the fourth circuit element relating the electrical charge with the magnetic flux[23],[24]. This element named *Memristor* as a combination of "Memory" and "Resistor". Prior to 2008 no successful implementation of this element was reported, this was mainly due to the fact that the memristive characteristic is only observable in Nano scale. In mid-2008, in HP research laboratory the first memristor was successfully realized [25]. Memristor has various applications such as implementing non-volatile RAM [26], spiking neural networks[27] and human learning algorithms[28],[29], digital circuits[30],[31], programmable analogue circuits[32],[33],[34],[35], pattern recognition and signal processing[36]. Because of the powerful implementation capabilities, low power consumption and stability of stored data even without power, memristor has received considerable attention[37]. Fig. 6 shows the relationships between different variables in electrical circuits.

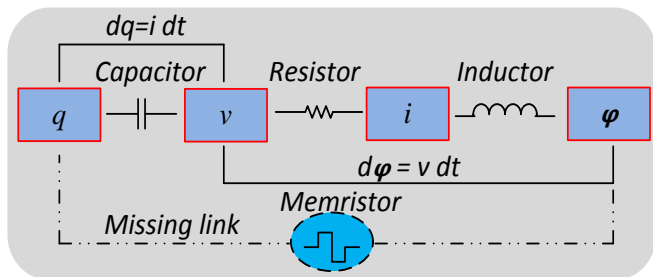

Fig. 6.  Four fundamental elements of electrical circuits: resistor, capacitor, inducer and memristor. Memristor is the fourth fundamental element and the missing link which relates electrical charge with magnetic flux. Theoretically, the value of a memristors remains unchanged if no current passes through it. Like capacitors, memristor does not require refreshment to preserve its value.

### A.  Physical structure and equations

Memristor is a passive two-terminal element which relates the electrical charge with magnetic flux as follows:

$$d\varphi = R_M dq \tag{13}$$

Equation (13) can also be written as follows, which indicates that the unit of memristance is Ohm.

$$R_M(q) = \frac{d\varphi/dt}{dq/dt} = \frac{v(t)}{i(t)} \tag{14}$$

Memristor can behave as a dynamic resistor whose resistance changes with respect to the voltage applied to or current passes through its terminals. If the characteristic of the memristor is considered linear, it behaves as a simple resistor with resistance $R_M$. Fig. 7 shows the physical and circuit model of the first memristor realized by HP [25]. As it can be seen from Fig. 7, memristor is comprised of a very thin layer of Titanium Dioxide ($TiO2$) with width $D$ sandwiched between two platinum ($Pt$) contacts. The semiconductor itself is comprised of a doped and an undoped regions. The width of doped region is $W(t)$ and the resistance of this region is lower than the other region. The variable width of the doped region makes the memristor a dynamic resistor.

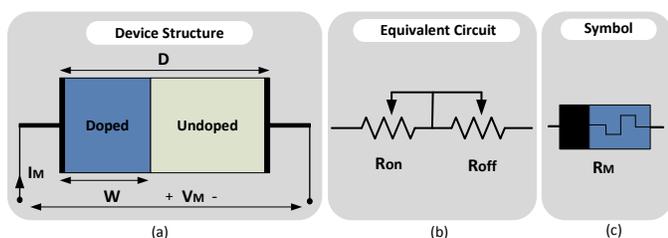

Fig. 7.  (a) The structure of a memristor is comprised of two regions; the doped region (low resistance $R_{on}$) and undoped region (high resistance $R_{off}$). By applying voltage with appropriate polarity and amplitude the border between these regions displaces. (b) Circuit equivalent of memristor. (c) memristor symbol.

By applying voltage to the two terminals of a memristor, the border of two regions displaces and this causes the total resistance to change. The resistance value changes between two extremes; $R_{on}$ and $R_{off}$. When the applied voltage amplitude is so that the doped region extends to the full width $D$ ($w \to D$) the memristance tends to reach its minimum value ($R_{on}$) and inversely when the applied voltage amplitude makes the undoped region extends to the full width $D$ ($w \to 0$) the memristance tends to reach its maximum value ($R_{off}$). In the mathematical model of memristor proposed in [25], as in (15) and (16), the electrical field in the memristor is assumed to be uniform.

$$w(t) = w_0 + \frac{\mu_v R_{on}}{D} q(t) \tag{15}$$

$$R_M(w) = R_{on}\frac{w}{D} + R_{off}(1 - \frac{w}{D}) \tag{16}$$

Where $w_0$ is the initial width of doped region $w$, $\mu_v$ is the average ion mobility and $q(t)$ is the net electrical charge passing through the element. These equations also reveal that the passing electrical current in one direction increases the memristance; whereas, the opposite direction of current decreases the memristance and if no current passes through the element, the memristance remains constant and behaves like a simple resistor. Thus, polarity and amplitude of the voltage signal as well as the duration of this signal are the main parameters affecting the memristance. In order to read the value of a memristor, it is sufficient to apply a current signal with amplitude less than a threshold for a short period of time and read the voltage over the terminals.

Various algorithms and computational frameworks for simulating the computational capabilities of the neural system of living organisms have been proposed. Almost all these frameworks suffer from a major drawback that is the lack of compatibility between hardware and the nature of the problems in hand such as implementation of Neuromorphic systems on FPGAs [38]. The main focus, though, is on hardware implementation which is not efficient that makes the large scale implementation of these algorithms infeasible. This problem has been resolved to some extent since the realization of memristor which complies with the synaptic behavior of a biological neuron and can be implemented in small size. Numerous studies have been shifted to this kind of implementations. In the next section the implementation of the IDS plane on the memristor crossbar structure will be presented.

### B.  Memristor crossbar implementation of IDS plane

A memristor crossbar is comprised of a series of horizontal wires passing over vertical ones, and at each intersection, a memristor is connected so that by applying proper voltage over any pair of vertical and horizontal wires, the memristor at that intersection can be accessed. In fact, the memristor crossbar is analogous to an array of analogue memory cells. The analogue values are stored as memristance of memristors.

The advantages of such an implementation are its nanoscale implementation and low power consumption.

As discussed in previous sections, the IDS operator in ALM requires spreading ink drops on 2-D memory planes. Because of the close resemblance between memristor crossbar structure and IDS planes, this structure has been employed to implement IDS operator. Fig. 8 shows such an implementation which was proposed in [21] and includes Memory Unit and Computation Unit as well. In this structure each memristor plays the role of a pixel in a 2-D IDS plane and its memristance, which can be set by applying proper voltage, is set to be the value of the corresponding pixel (pixel darkness).

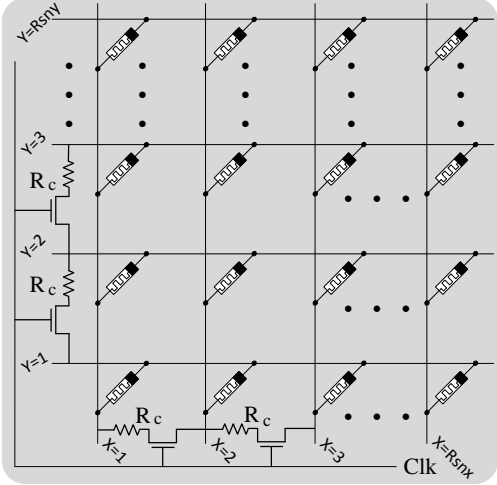

Fig. 8. The implementation of IDS plane on memristor crossbar structure. This structure was first proposed by Merrikh-bayat in [21] which is designed for spreading ink drops on an IDS plane with resolution $Rsn_x * Rsn_y$. Resistors with constant resistance $R_c$ are employed to perform the spreading of ink drop (blurring). In the training phase when the CLK signal is active, resistors $R_c$ become part of the circuit.

Memristor crossbar structures are able to store the data in analogue form and thus have higher capacity compared to digital memory structures. Furthermore, these structures preserve data without energy consumption, and in comparison with digital memories, faster reading and writing are possible. In order to model a system with $N$ inputs with ALM, at least $N$ 2-D IDS planes or memristor crossbar structures are required. In addition to memory units, a number of computational circuits are required to extract *Narrow Path* and *Spread*. However, by proposing a novel view to inference step of ALM and compromising the accuracy, Esmaili in [22] proposed a novel approach which relaxed the large computational hardware requirement, yet the large number of circuit elements was apparent. In the next section, by incorporating a novel perspective on ink drop spread on IDS planes, we fully describe our contribution which aims at reducing the complexity of computation and hardware implementation.

## IV. PROPOSED ALGORITHM

One of the main challenges in the ALM algorithm, in particular IDS unit, is its hardware implementation and large amount of memory required to store the information of IDS planes. This becomes even worse when partitioning the inputs domains is desirable. In the original IDS unit, the size of the required memory is as large as the whole IDS plane grids. In [5] an alternative hardware implementation of IDS unit was proposed that employs two memory vectors where any pair of Euclidian adjacent points are replaced by their mean. In this approach the size of the required memory space diminishes seven times. A digital parallel hardware implementation was also proposed in [19] that suffers from digital problems such as overflow and limited precision. Therefore, seeking an alternative approach to decrease the complexity of IDS computations is necessary.

The most critical step in proposing a more efficient alternative IDS algorithm is to find a proper mapping in order to represent the data space. The new mapping should deal with computational complexity, complicated equations and hardware limitation to obtain the highest possible precision. In addition, new mapping should be suitable for real-time applications. Having smaller and denser structure compared to other circuit elements, memristor implementation also operates with lower power consumption. All these advantages make memristor implementation a suitable structure for our purpose. In the proposed algorithm three memory vectors are employed to describe and store all valuable features of an IDS plane, two of which are responsible for storing the lower-bound and the upper-bound of the output. And the third one stores the *Narrow Path*. In the proposed algorithm three separate units are required namely Storage, Learning and Features Extraction units. Moreover, a high level controller is required to generate electric pulses used in learning phase and reading the memristance of memristors. Inputs domain determination and partitioning are done by the human operator and then appropriate electric pulses are generated in Chip Programmer to perform the learning phase in the Memristor Chip. Fig. 9 shows the systematic structure of the new IDS unit (FAST IDS). These units will be discussed in more details in following subsections.

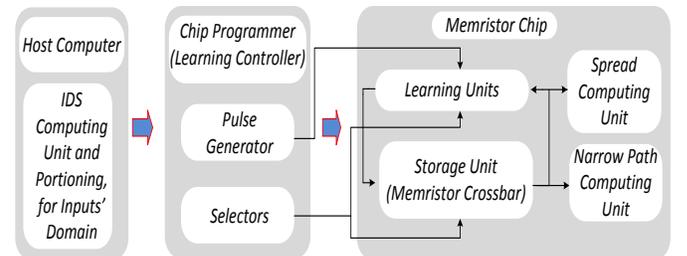

Fig. 9. The systematic implementation of the proposed IDS unit (FAST IDS). All the primary computations are done in the host computer and then by generating appropriate electric pulses in the Chip Programmer, the learning process is performed on the proposed hardware.

Like all learning algorithms, there are learning and test phases, which will be discussed in following subsections.

### A. Initialization and Learning Algorithm

Let $S = \{(x_1, y_1), (x_2, y_2), ..., (x_N, y_N)\}$, $x = (x_1, x_2, ..., x_D)^T$ where $N$ and $D$ are the number of samples and the number of independent variables respectively. In our proposed algorithm, there are three assumed vectors that are denoted by $c_{LB}, c_{UB}, c_{NP}$ that describe IDS planes. All these vectors are $Rsn_x$ long. $c_{LB}$ and $c_{UB}$ denote the output lower-bound and upper-bound respectively. $c_{NP}$ is a coefficient of each output sample and is equivalent to *Narrow Path* in original IDS and eventually converges to it. The difference between. $c_{LB}$ and $c_{UB}$ acts like the *Spread* in the original IDS and specifies the degree of certainty around each input point. Thus, the initial values of describing vectors are as follows:

$$c_{LB}(initial) = 0 \qquad (17)$$

$$c_{UB}(initial) = Rsn_y \qquad (18)$$

$$c_{NP}(initial) = \frac{Rsn_y}{2} \qquad (19)$$

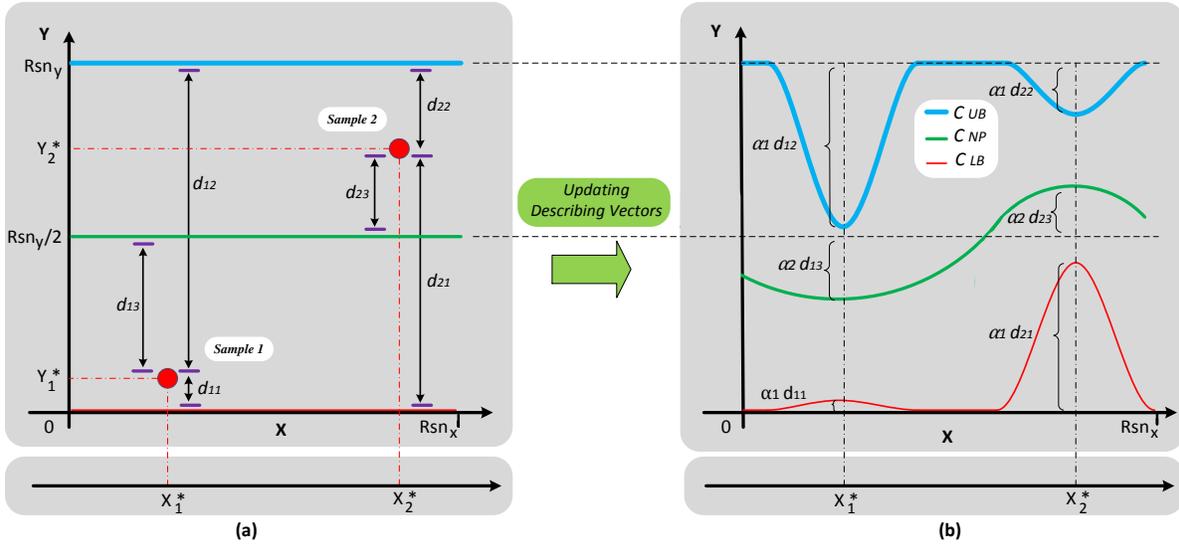

Fig. 10.  An example of updating describing vectors. (a) Two training samples are observed $(x_{\text{train}1}, y_1) = (8,8)$, $(x_{\text{train}2}, y_2) = (4,3)$. The input and output domains are determined as $[x_{\min}, x_{\max}] = [1,10]$, $[y_{\min}, y_{\max}] = [1,10]$. With resolution $Rsn_x = Rsn_y = 256$ the quantized samples are $(x_{\text{new}1}, y_{\text{new}1}) = (200,200)$, $(x_{\text{new}2}, y_{\text{new}2}) = (86,57)$. (b) Updating describing vectors with regard to their vertical distance to the new observed sample. In this example parameters are as follows $\alpha_1 = 0.6$, $\alpha_2 = 0.5$, $\sigma = 15$ .

The learning algorithm is as follows. When a new training sample is observed, values stored in three describing vectors are updated with regard to their distance to the output value of the observed sample. As mentioned before, the output value of each sample is quantized between 0 and $Rsn_x$.

For each observed sample, its output distance to all describing vectors are calculated. These vectors are then updated locally. If $q(x)$ is a point on the x axis and $(x_s, y_s)$ is the training sample, updating rules of describing vectors are as follows:

$$Q(x) = \{q(x) | x \in X_i\} \qquad , -r \le u \le r \qquad (20)$$

$$c_{LB}(x_s + u) = c_{LB}(x_s + u) + \alpha_1 * [y_s - c_{LB}(x_s)] \\ * g(u) \qquad (21)$$

$$c_{UB}(x_s + u) = c_{UB}(x_s + u) + \alpha_1 * [y_s - c_{UB}(x_s)] \\ * g(u) \qquad (22)$$

$$c_{NP}(x_s + u) = c_{NP}(x_s + u) + \alpha_2 * [y_s - c_{NP}(x_s)] \\ * g(u) \qquad (23)$$

and for $g(u)$ we will have:

$$g(u) = \exp(-\frac{u^2}{2\sigma^2}) \qquad (24)$$

Where $g$ is a Gaussian function with variance $\sigma$. There are three parameters used in updating rules $\alpha_1$, $\alpha_2$ and $\sigma$. Learning rates $\alpha_1$, $\alpha_2$ are employed to specify the extent to which describing vectors are curved toward the observed samples. These two parameters also affect the convergence of describing vectors toward training samples. Variance $\sigma$ specifies the local neighborhood of an observed sample that updating rules are allowed to expand their impacts. This parameter is set to be large for sparse datasets to cover more regions and is set to be small in dense datasets. The parameters setting plays an important role in performance as well as convergence speed. In sparse datasets to increase the output precision, these parameters can be chosen to be large. With small parameters, multiple epochs learning, like artificial neural networks, is also available in this algorithm that can result in higher precision. These parameters can be set either by trial and error or by optimization algorithms such as Genetic Algorithm. Fig. 10 shows an example of updating describing vectors in a scenario where two training samples are observed.

The proposed algorithm provides a novel description of the IDS plane by introducing three vectors. Despite original IDS algorithm that requires a $Rsn_x \times Rsn_y$ memory matrix to store the information of each IDS plane, in the proposed algorithm three vectors with length $Rsn_x$ are sufficient. This results in considerable decrease in memory usage. In the test phase, it is sufficient to read the associated elements of three describing vectors and then feed these values to the inference engine of ALM algorithm. In this section learning phase and convergence of the algorithm is presented through an illustrative example. As it was discussed in this section, the ink drop spreading process has been removed and thus its computational complexities have been avoided. In the rest of paper, we use the term "Fast ALM" in order to refer to ALM algorithm whose IDS unit is replaced by Fast IDS unit. The next section deals with Memristor-CMOS implementation of our proposed algorithm and advantages of our proposed algorithm will be presented by comparing its circuit complexity with that of traditional implementations.

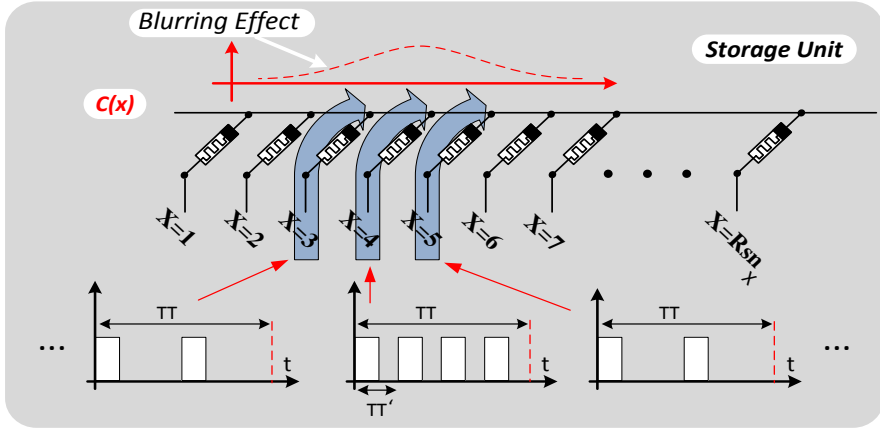

Fig. 11. The Proposed memristor crossbar structure for storing features of the IDS plane. In this example, a typical sample in x=4 is assumed for first row of the memristor. Squared voltage signal with appropriate amplitude and width is applied to this input and as it moves further form this input, the effective width of squared voltage signal is decreased to model the blurring effect (the effect of the function g(u) in (24)).

## V.    HARDWARE OF THE PROPOSED ALGORITHM

In this section the hardware implementation of our proposed algorithm is presented. The proposed hardware has units for updating IDS planes (learning phase) as well as Feature Extraction Unit (inference or test phase). As mentioned in previous sections, memristor is a memory element that is suitable for storing analogue values. Fig. 11 shows a memristor crossbar structure with two rows of memristors for storing two describing vectors $c_{LB}$ and $c_{UB}$. Despite hardware implementation proposed in [21] (Fig. 8), there is no need for $R_C$ resistors. Like hardware implementation proposed in [22], in order to provide symmetric blurring effect, there is no need for resistive ladder structure that leads to some complexities in hardware implementation. In our implementation, in order to change the values of memristors for each input point it is sufficient to apply voltage with amplitude proportional to the desirable memristance change. Based on the user-defined variance, adjacent memristors also take effect by applying this voltage signal.

Learning algorithms are comprised of two main phases; learning phase and test phase. The proposed hardware should have circuits to perform learning and test phases as well. In the next subsection, circuit required to perform learning phase is presented.

### A.   Initialization

In learning algorithms, the first step is to learn from the training set. As it can be seen from Fig. 12 the proposed circuit is symmetric where the Fig. 12(a-1) and Fig. 12(a-2) circuits store the upper-bound and the lower-bound of the output respectively. Storage Unit also has a demultiplexer with address line for choosing input $X$ and its neighbors (m memristors as shown in Fig. 12) to apply the blurring effect. Fig. 12(b-1) and Fig. 12(b-2) units are for learning the upper-bound and the lower-bound of the output respectively. Each of these units has five stages including Op-amp amplifier, MOSFET switches, capacitor, training signals $v_{train}$, $v_{bias}$ and $v_{read}$ and Connector Block. Unit Fig. 12(c-1) is a differential Op-amp amplifier that extracts the degree of certainty or membership function by measuring the difference between the lower-bound and the upper-bound.

Like other learning algorithms, parameter initialization is the first step of our proposed algorithm; thus, the initial memristance of all memristors is set to be between the maximum and minimum resistance $0 < R_{on} \le R_{M_i} \le R_{off}$. In what follows, initial resistance of each memristive memory array is computed. Each memristor in this structure is connected to the negative terminal of an Op-Amp. This Op-Amp computes the weighted sum of input signals. This weight or gain is equal to the negative ratio of feedback resistor of the Op-Amp to memristor impedance. The feedback resistance in the first stage is equal to $R_{F1}$, therefore the gain of this stage is $R_{F1}/R_{M_i}$. If the output voltage of this stage is $g_i$ we have:

$$g_{UB_i} = v_{Read} * \frac{R_{F1}}{R_{M_{Ui}}} \ , \ \ g_{LB_i} = v_{Read} * \frac{R_{F1}}{R_{M_{Li}}} \tag{25}$$

and for the second stage we have:

$$z_{UB_i} = v_{Bias} - v_{Read} * \frac{R_{F1}}{R_{M_{Ui}}} \tag{26}$$

$$z_{LB_i} = v_{Bias} - v_{Read} * \frac{R_{F1}}{R_{M_{Li}}} \tag{27}$$

Initially, the output node $SP$ should be set to its maximum value, which corresponds to the lowest degree of certainty. This entails that node $z_{UB_i}$ should have the maximum voltage $v_{Read}$ and node $z_{LB_i}$ should have the minimum voltage, zero. If in the Fig. 12(b-1) circuit $R_{F1} = R_{on}$ and the initial value of all $R_{MU_i}$ is assumed to be equal to $R_{off}$, then because

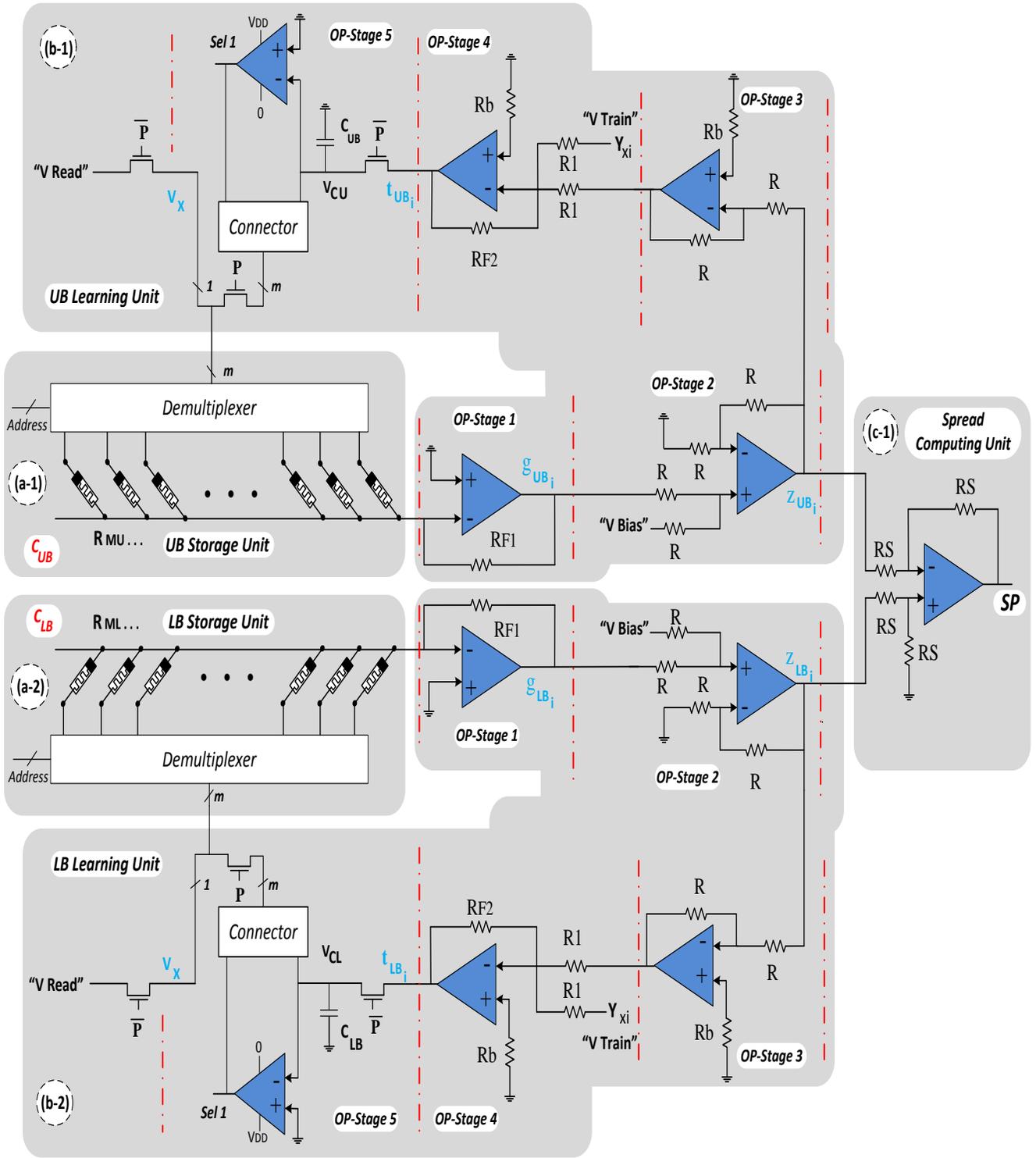

Fig. 12. The proposed hardware implementation for storing, learning and spread computing processes. Circuit (a) of the proposed hardware is responsible for storing the upper-bound and the lower-bound of the input and circuit (b) performs the process of learning and storing in memristors. Circuit (c) is also known as Spread Computing Unit that measures the spread of an input point. In this circuit parameter $\alpha_1$ in (21) and (22) equals to $\frac{R_{F_2}}{R_1}$.

$\frac{R_{on}}{R_{off}} \ll 1$, the initial voltage of node $z_{UB_i}$ is equal to $v_{Bias}$. By decreasing the resistance of $R_{MU_i}$ toward $R_{on}$, if $v_{Bias} = v_{Read}$, then the initial voltage of node $z_{UB_i} = v_{Bias} - v_{Read}$ becomes zero. In Fig. 12(b-2) if $R_{F1} = R_{on}$ and if the initial value of all $R_{MU_i}$ is assumed to be equal to $R_{on}$, then the initial voltage of node $z_{LB_i}$ is equal to zero. By increasing the resistance of $R_{ML_i}$ toward $R_{off}$, then the initial voltage of node becomes $z_{LB_i} = v_{Bias}$. Therefore, all memristors of the first and second rows are initialized to $R_{off}$ and $R_{on}$ respectively. Note the polarity and the connection of memristors in Fig. 12.

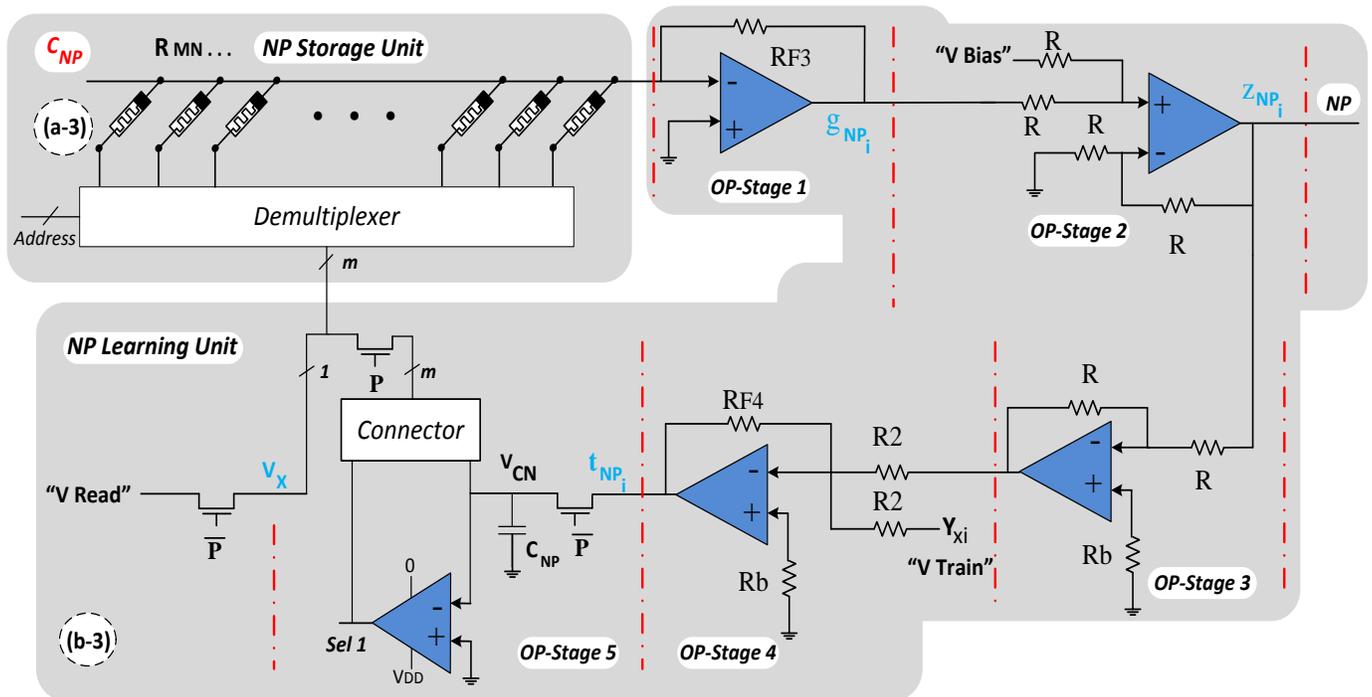

Fig. 13. The designed circuit for learning, storing and extracting *Narrow Path*. Like the Spread Computing Circuit, this circuit is comprised of Storage Unit (a) and Learning Unit (b). The voltage of output node $z_i$ is proportional to voltage node $NP$. This circuit has its own learning rate $\alpha_2$, which is specified by ratio $\frac{R_{F4}}{R_2}$ as defined in (23).

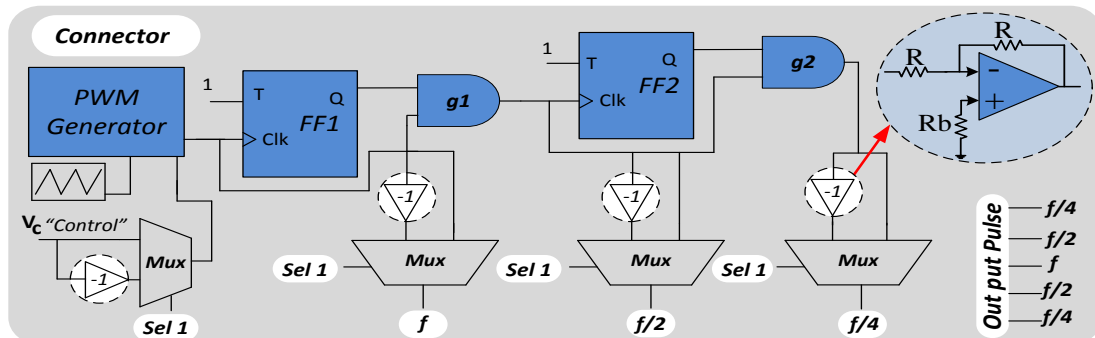

Fig. 14. Connector Circuit. In this circuit a PWM generator receives the period $T$ from a triangular signal with fixed frequency and then generates a squared signal f whose amplitude and duty-cycle is controlled by the voltage of a capacitor. Signal $f$ and the output of the first Flip-Flop $FF1$ are then passed through an AND gate and generate signal $f/2$ whose overall high time is half that of signal f (for $k$ period of signal $f$ (in this figure k is 4). Signal $f/2$ and the output of the second Flip-Flop $FF2$ are then passed through an AND gate and generate signal $f/4$ whose overall high time is a quarter that of signal $f$. According to the variance defined by the user, more Flip-Flop and gate stages can be added to generate appropriate signals to affect more neighbors of the targeted memristor (Blurring effect). For the cases when the voltage of the capacitor is negative, the negating circuit is included to generate signals with negative amplitude. The negative signals can be chosen by selecting line sel1.

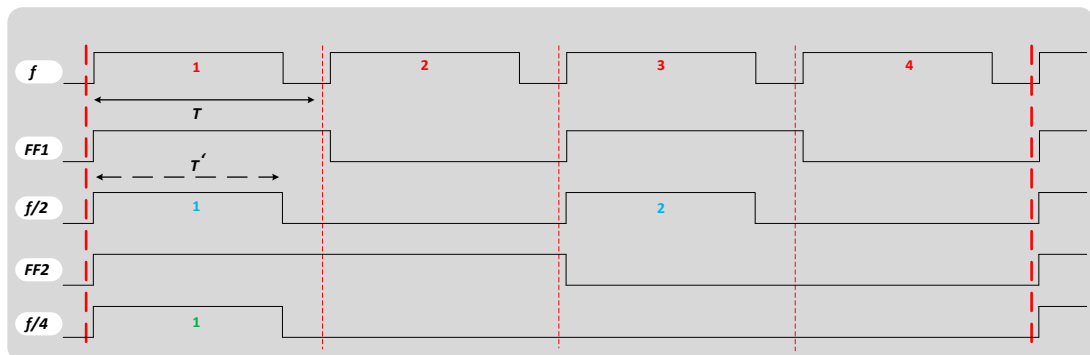

Fig. 15. Generated signals in the Connector Circuit for positive voltage of the capacitor. The period of the generated PWM signal is T and its duty cycle is controlled by a capacitor voltage. The higher the voltage of the capacitor, the wider the duty cycle of generated PWM signal. These signals have fixed frequency and their width and the amplitude should be sufficient to change the memristance of memristors.

In Fig. 13, the *Narrow Path* Extraction Circuit is shown. In the proposed circuit, the voltage of the output node $NP$ should be initialized to the half of the resolution of the IDS plane. This entails that the voltage of the node $z_{NP_i}$ should be equal to $v_{Read}/2$. In the Fig. 13(b-3), if $R_{F3} = R_{on}$, $v_{Bias} = v_{Read}$ and $z_{NP_i} = v_{Bias} - v_{Read} * \frac{R_{F3}}{R_{MN_i}}$ and we assume that the initial resistance of $R_{MN_i}$ is equal to $2R_{on}$, then the voltage of node $z_{NP_i}$ is equal to $v_{Read}/2$. Increasing or decreasing the resistance of $R_{MN_i}$ increases or decreases the voltage of node $z_{NP_i}$ respectively. Fig. 14 shows the Connector Circuit which is comprised of Flip-Flops, AND gates, Triangular signal generator, Multiplexers and negating Op-Amp circuits. The latter circuit receives a triangular signal with constant frequency and amplitude, which is stored in a capacitor, and then generates appropriate squared signals required in the learning phase. The width of squared signals for central memristor $x_i$ and m signals of its neighbors (for simplicity $m = 5$) are adjusted in this circuit.

Fig. 15 shows the signals generated in the Connector Circuit. These signals are selected based on the polarity of the capacitor and are applied to memristors in the learning phase.

### B. Learning Phase

The learning phase in the proposed hardware is as follows. When a new training sample is observed, its distance to all elements of three describing vectors are calculated and then multiplied by learning rates $\alpha_1$ and $\alpha_2$ to update three describing vectors, as shown in Fig. 10. The process involved in reading memristors values and updating them is the same for each describing vector. In order to read memristance of a memristor, signal $\bar{P}$ is activated and as a result, the MOSFET switch is turned on. By applying appropriate voltage signal $v_{read}$ to the targeted memristor by address line of the demultiplexer, the reading process will be started. Voltage $v_{read}$ reaches the output node $g_{UB_i}$ with amplification gain $\frac{R_{F1}}{R_{MU_i}}$. The output of each Op-Amp stage is as follows:

$$v_{out-op3} = -z_{UB_i} \quad , \quad \frac{R_{F2}}{R_1} = \alpha \qquad (28)$$

For voltage y proportional to each sample, proportional voltage $v_{read}$ should be applied, which is computed as follows:

$$v_{CU} = t_{UB_i} = -z_{UB_i} * \frac{R_{F2}}{R_2} + v_{Train} * \frac{R_{F2}}{R_1} \qquad (29)$$

$$v_{out-op3} = -v_{Read} * \left(1 - \frac{R_{F1}}{R_{MU_i}}\right) * \frac{R_{F2}}{R_2} + v_{Train} \\ * \frac{R_{F2}}{R_1} \qquad (30)$$

$$t_{UB_i} = \alpha(v_{Train} - v_{Read} * (1 - \frac{R_{F1}}{R_{MU_i}})) \qquad (31)$$

If the MOSFET switch $\bar{P}$ is activated, the capacitor $v_{CU}$ holds the voltage $t_{UB_i}$. In the next step, by activating the MOSFET switch $P$, the stored voltage in the capacitor is transferred to Connector Circuit. Based on user-defined variance $\sigma$, which defines the number of neighbor memristors to be involved in the learning process when the targeted memristor is chosen to be updated, the appropriate output signals are applied to these memristors. The Voltage $t_{UB_i}$ obtained in (31) is equivalent to (21), where $v_{train}$ is proportional to the output value of training sample $y_s$ and it is proportional to $v_{read}$ which is itself proportional to the value stored in the memristor array $c(x_s)$. This equivalency is shown in (33).

$$v_{CU} = \alpha * (v_{Train} - v_{Read} * (1 - \frac{R_{F1}}{R_{MU_i}})) \overset{Eq.\ 21}{\approx} \alpha \\ * [y_s - c_{UB}(x_s)] \qquad (33)$$

As shown in (33), the output voltage stored in the capacitor is proportional to the distance. Thus the proposed circuit implements the algorithm with acceptable approximation (for two other vectors we do the same as (33)).

### C. Modeling or Test Phase

In the previous subsection the learning phase of the proposed hardware was presented. In this subsection, the required hardware for the test phase, when it should model previously unseen test samples, is proposed. In order to perform the inference in the proposed algorithm, it is first required that the *Narrow Path* and *Spread* values in the coordination of the test sample be extracted from three describing vectors of each IDS plane. The proposed circuit for extracting *Spread* is shown in Fig. 12(c-1). We have:

$$SP_{xi} = z_{UB_i} - z_{LB_i} \qquad (34)$$

The proposed hardware acts like a differential amplifier that amplifies the difference between $z_{UB_i}$ and $z_{LB_i}$. In Fig. 13(b-2) the output voltage $z_{L_i}$ is proportional to *Narrow Path* for given coordination of the test sample. These values are used in the inference engine of ALM. In the test phase, for each input value x = x_i, *Narrow Path* and *Spread* are available for any time. For each input x, only the associated features are extracted and the computation and storage of all values of *Narrow Path* and *Spread* are not required.

The proposed circuit performs well only when appropriate signals are applied to the memristor circuit, so the value of memristor changes in the desirable direction with desirable amplitude. Therefore, in order to read each memristor, it is sufficient to apply voltage $v_{read}$ to the column of the memristor associated with test sample and apply appropriate squared signal to its row and read the output voltage. The frequency of such a signal should be high and its width should be so negligible that by applying it to the memristor, it acts like a resistor and its memristance can be read without any change. In order to write a new value in a resistor the appropriate squared signal should have two main characteristics; the amplitude of such a signal should be greater than the threshold of memristor and it should be wide enough to change the memristance.

For higher precision more quantization levels are required. This is achieved by adding more memristor to the Storage circuit. Table I Compares the circuit complexity of our proposed algorithm (Fast ALM) and ALM[21].

TABLE I
THE COMPARISON OF CIRCUIT COMPLEXITY OF OUR PROPOSED ALGORITHM (FAST ALM) AND ALM[21]. THE COMPARISON IS DRAWN FOR A SINGLE-INPUT SINGLE-OUTPUT SYSTEM WITHOUT PARTITIONING THE INPUT DOMAIN (ONE IDS PLANE).

| Elements | ALM in [21] | Fast ALM |
|---|---|---|
| Opamp | $2*R_{snx}$ | 28 |
| Memristor | $R_{snx}*R_{sny}$ | $3*R_{snx}$ |

Unlike ALM, in Fast ALM the number of Op-Amps is independent of the resolution of the IDS plane. Table I Also reveals that in terms of size and power consumption our proposed hardware is considerably more efficient than the hardware proposed in [21]. Furthermore, the proposed hardware has low circuit complexity and faster process time without compromising the precision.

Another advantage of our proposed hardware compared to traditional hardware is that the need for so many multipliers and adders is satisfied to a great extent. It should be mentioned that in this comparison only shared circuit elements in both hardware implementations have been taken into account and those circuit elements such as Flip-Flops, AND gates, etc. of each specific hardware are excluded.

## VI. SIMULATION

In this section the functionality and performance of our proposed algorithm is evaluated on various applications. Function approximation and classification are two problems considered in this section. All simulations are performed in MATLAB 2013 on Core i5 processor, 2.4GHz, 4GB RAM. The quantization levels in all simulations are considered $Rsn_x = 256 * Rsn_y = 256$.

### A. Function Approximation

In this section, the performance of Fast ALM algorithm and its hardware implementation is evaluated on function approximation problem. Two 2-input single-output functions are considered as follows.

$$F_1 = (1 + x_1^{-2} + x_2^{-1.5})^2 \qquad , 1 \leq x_1, x_2 \leq 10 \qquad (35)$$

$$F_2 = \sqrt{2 * \left(\frac{sinx_1}{x_1}\right)^2 + 3 * \left(\frac{sinx_2}{x_2}\right)^2}, 1 \leq x_1, x_2 \leq 10 \qquad (36)$$

Fig. 16 and 17 show two functions as defined in (35) and (36). In this proposed hardware the period of the input triangular signal in PWM generator is 10milliseconds ($T = 10ms$). The output of PWM generator is 3 volt amplitude and 80 percent duty cycle ($T' = 10ms$). Simulation have been conducted in HSPICE software and memristor's SPICE model has been obtained from[39]. Finally memristor's parameters were set as follows:$D = 10e - 9, \mu = 1e - 14, R_{off} = 200\ K\Omega$ and $R_{on} = 2K\Omega$.

TABLE II
THE COMPARISON OF ALM AND FAST ALM BASED ON FVU METRIC. THE RESULTS ARE THE AVERAGE OF 100 RUNS. IN THIS EXPERIMENT, THE VARIANCE FOR 2-D INK DROP SPREADING IN ALM AND 1-D GAUSSIAN FUNCTION IN FAST ALM (IN (24)) IS THE SAME AND EQUAL TO 15. IN FAST ALM, $\alpha_1 = 0.01, \alpha_2 = 0.95$. THE TRAINING PHASE IS PERFORMED WITH 2500 TRAINING SAMPLES. THE PARTITIONING POINTS ON INPUTS DOMAIN ARE THE SAME FOR BOTH ALGORITHMS. THE PARTITIONING POINTS ARE CHOSEN RANDOMLY WITHOUT ANY OPTIMIZATION TOOL.

| | | | Number of training samples | | | | | |
|---|---|---|---|---|---|---|---|---|
| Number of partitions | | | 225 | | 550 | | 1000 | |
| | | Function | Fast ALM | ALM | Fast ALM | ALM | Fast ALM | ALM |
| X1 | X2 | | | | | | | |
| 2 | 2 | F1 | 0.1579 | 0.1269 | 0.1429 | 0.1311 | 0.1828 | 0.1342 |
| | | F2 | 0.0929 | 0.0942 | 0.0663 | 0.0862 | 0.0723 | 0.0821 |
| 4 | 4 | F1 | 0.1089 | 0.1374 | 0.0601 | 0.0790 | 0.0562 | 0.0744 |
| | | F2 | 0.0772 | 0.0823 | 0.0588 | 0.0698 | 0.0412 | 0.0669 |
| 8 | 8 | F1 | 0.1475 | 0.1166 | 0.0840 | 0.0660 | 0.0867 | 0.0978 |
| | | F2 | 0.0827 | 0.0791 | 0.0441 | 0.0584 | 0.0534 | 0.0501 |
| 10 | 10 | F1 | 0.1106 | 0.1100 | 0.0814 | 0.1186 | 0.0493 | 0.0801 |
| | | F2 | 0.0790 | 0.0781 | 0.0437 | 0.0555 | 0.0642 | 0.0570 |

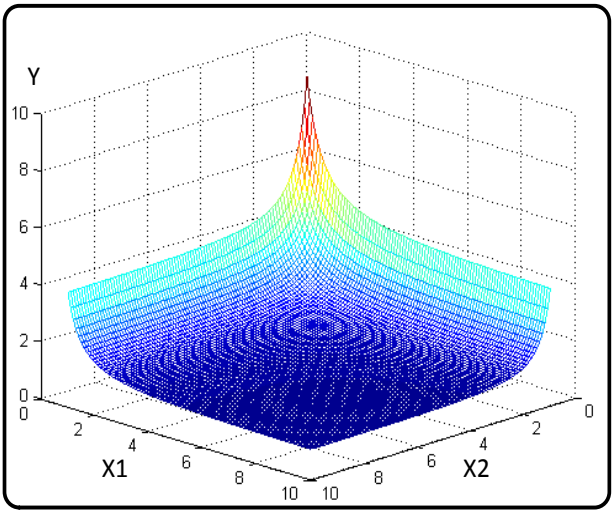

Fig. 16  Function $F_1$ as defined in (35).

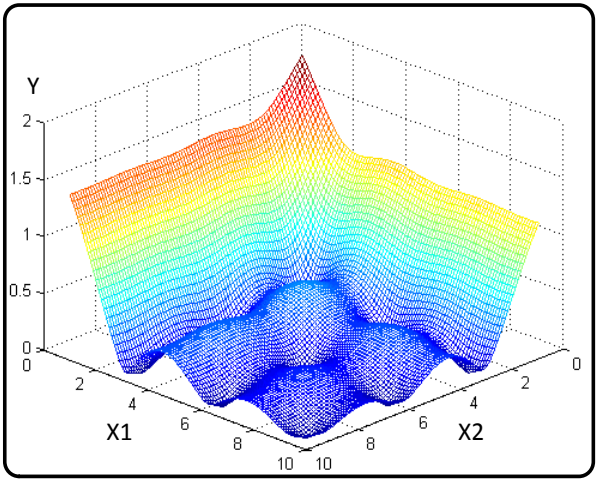

Fig. 17  Function $F_2$ as defined in (36).

To evaluate the precision and approximation error, a metric employed by Sugeno in [8] was used, which is called Fraction of Variance Unexplained (FVU). This metric is directly related to MSE so that if FVU is zero, MSE will be zero too. In (37), the higher the precision, the closer FVU is to zero.

$$FVU = \frac{\sum_{i=1}^{k}(\hat{y}(x_i) - y(x_i))^2}{\sum_{i=1}^{k}(y(x_i) - \bar{y})^2} \tag{37}$$

$$\bar{y} = \left(\frac{1}{k}\right)\sum_{i=1}^{k} y(x_i) \tag{38}$$

In this equation, $\bar{y}$ is the output of the model and $K$ is the number of training samples. The simulation is performed with different number of training samples and different number of partitions. Because of the random choice of training samples, the results shown in Table II are the average output of 100 runs. In the simulation, first Fast ALM splits the two-input single-output system to two single-input single-output subsystems and the behavior of each subsystem on IDS plane is captured in describing vectors. According to desirable precision, partitioning the inputs domain can be done.

As it can be seen from the Table II, the proposed algorithm performs well in approximating complex functions. Like ALM, in Fast ALM when the training set is small (low knowledge) or IDS planes become sparse as a result of over partitioning, the approximation error rises. Whereas, when the training set is large enough, partitioning the input domain increases the precision. To illustrate more, Fig. 18 shows the learned pattern and convergence of the *Narrow Path* and *Spread* for a typical IDS plane in IDS and Fast IDS.

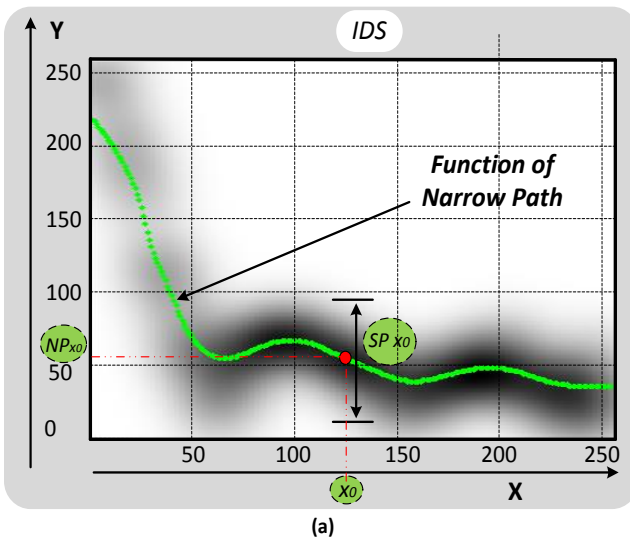

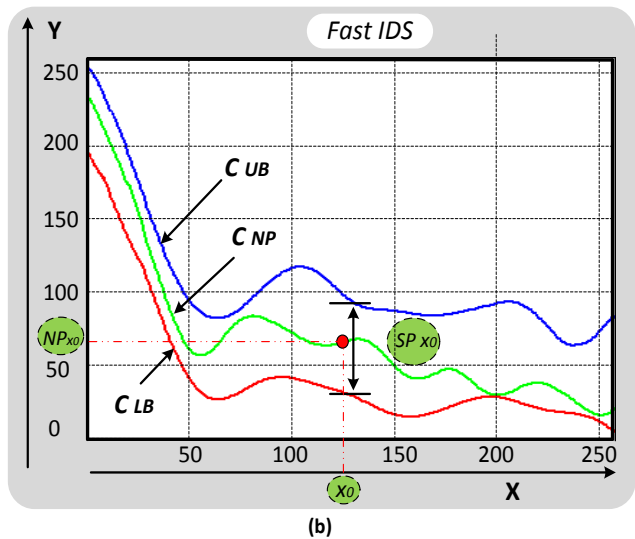

Fig. 18  The Comparison of the result of the proposed IDS (Fast IDS) and IDS. *Narrow Path* and *Spread* on an IDS plane associated with function $F_2$.  As it can be seen, the result of Fast IDS is close to that of IDS which means the mapping by describing vectors has been done perfectly. The variance $\sigma$ for both algorithms is the same and equal to 10. For Fast IDS $\alpha_1 = 0.65, \alpha_2 = 0.95$. *Narrow Path* and *Spread* for typical test sample $x_0$ is shown.

Fig. 19 and Fig. 20 show the approximated function by Fast ALM.

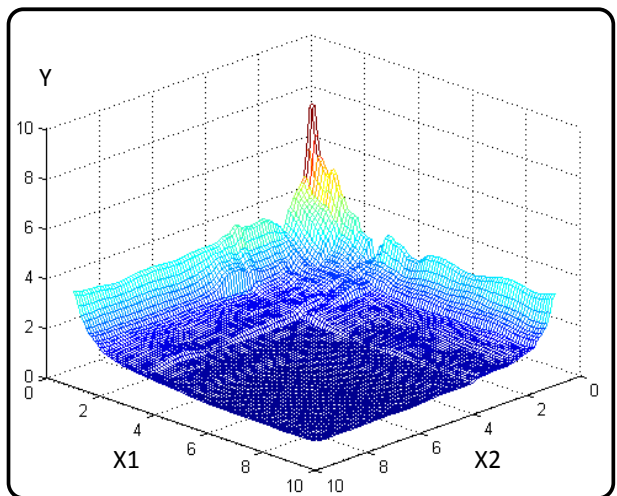

Fig. 19 The result of approximating function $F_1$ (in (35)) with 2500 training samples. 11 partitions are applied. Parameters are as follows. $\sigma = 12, \alpha_1 = 0.02, \alpha_2 = 0.92$ . FVU=0.0419 for this result.

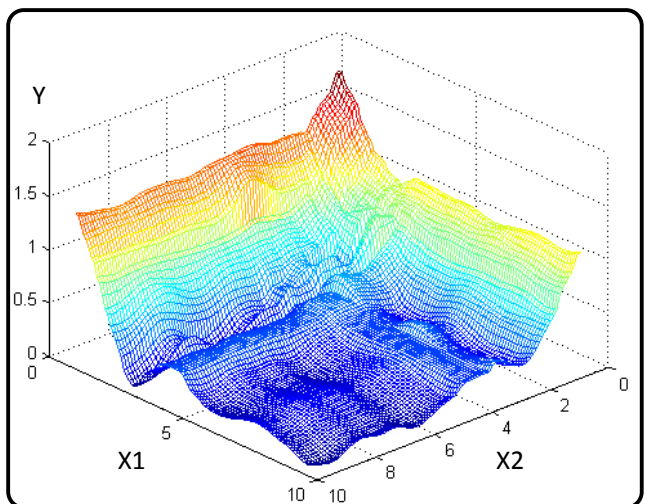

Fig. 20 The result of approximating function $F_1$ (in (35)) with 2500 training samples. 11 partitions are applied. Parameters are as follows. $\sigma = 10, \alpha_1 = 0.01, \alpha_2 = 0.96$ . FVU=0.0327 for this result.

In order to evaluate the process time and the convergence speed of Fast ALM compared to ALM, function $F_2$ with 2500 training samples is considered. In this experiment for two partitions on each input domain, Fast ALM and ALM converge in 0.1367 and 2.5990 seconds respectively.

Increasing the partitions to four on each input domain results in 0.2475 and 5.2281 seconds respectively. Fig. 21 shows the convergence speed with respect to the number of training samples.

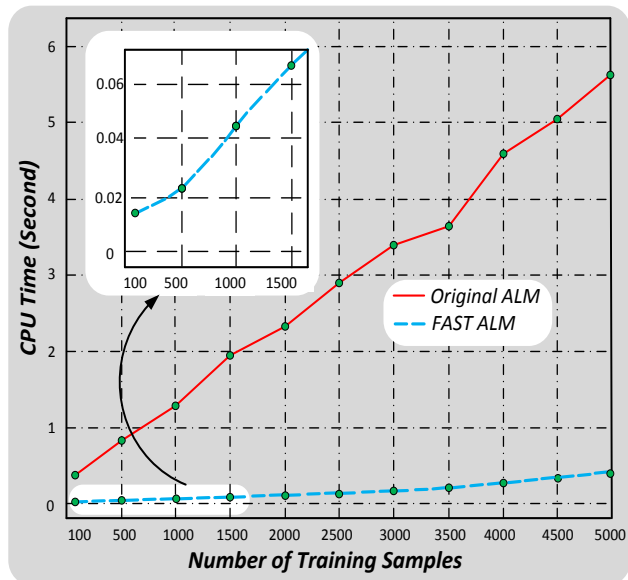

Fig. 21 The convergence speed of Fast ALM and ALM in approximating function $F_2$. The convergence speed of Fast ALM is considerably higher than that of ALM. In this experiment 8 partitions are considered and the convergence speed is plotted with respect to the number of training samples. Variance for both algorithms is the same and equal to 12. In Fast ALM $\alpha_1 = 0.01, \alpha_2 = 0.9$.

As it can be seen from simulations, Fast ALM achieves high precision in approximating functions. Next subsection examines the performance of FALM on classification problems.

### B.  Classification

In order to further examine the functionality of Fast ALM, its performance on classification problem is presented. To do so, two well-known classification problems called Two-Spiral [40] and Three-Centered-Ring are considered.

Two-Spiral dataset has two classes labelled 0 and 1. When no partitioning is performed, both algorithms classify randomly and do not show acceptable result; however, with appropriate partitioning, both algorithms can learn to correctly classify the dataset. Samples of this dataset are obtained from (39).

$$r = p(\theta + 2\pi n) + r_0 \qquad (39)$$

TABLE III
THE PERFORMANCE OF FAST ALM AND ALM IN CLASSIFYING TWO-SPIRAL DATASET WITH DIFFERENT PARTITIONING. THE RESULT IS THE AVERAGE OF 100 RUNS. PARTITIONING POINTS ARE THE SAME FOR BOTH ALGORITHMS. PARAMETERS ARE AS FOLLOWS. $\sigma = 4, \alpha_1 = 0.027, \alpha_2 = 0.23$.

| Classification Method | Test Set Size | Number of Partitions | | | | | | | |
|---|---|---|---|---|---|---|---|---|---|
| | | X1 | X2 | X1 | X2 | X1 | X2 | X1 | X2 |
| | | 4 | 4 | 5 | 5 | 6 | 6 | 6 | 8 |
| *ALM* | 600 | 0.83 | | 0.88 | | 0.91 | | 0.95 | |
| *FALM* | | 0.85 | | 0.87 | | 0.92 | | 0.94 | |

Where $r$ is radius, $\theta$ is angle in radian, $p$ is the radius of spiral ring, $r_0$ is the initial radius and $n$ is the number of spiral rounds. Fig. 22 shows Two-spiral dataset with 196 samples overall.

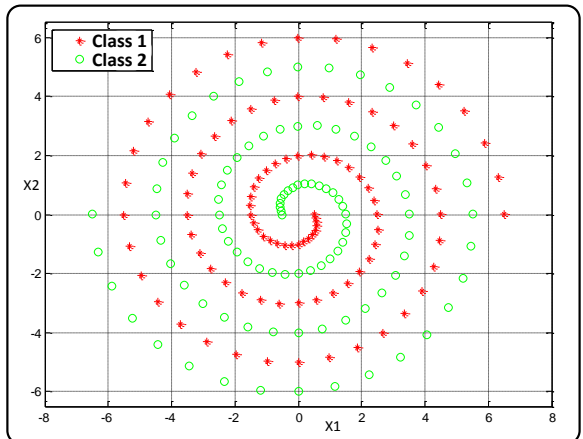

Fig. 22　Two-Spiral dataset in Cartesian coordinates. Each class has 97 samples. Parameters are as follows $p = 1/\pi$, $n = 3$, $r_0 = 0.5$.

To evaluate Fast ALM on 2-Spiral classification, first 400 training samples are introduced to both algorithms and their performances are evaluated in classifying 600 unseen test samples and the results shown in Table III.

As another example, the problem of Three-Centered-Ring is considered. Samples are positioned on three centered rings with different radiuses. Equation (40) defines the equations from which samples of each class are obtained. In this dataset three class labels 0, 1, 2 are considered.

$$
\begin{aligned}
Class\,1 &: x_1^2 + x_2^2 < R_1 \\
Class\,2 &: R_1 < x_1^2 + x_2^2 < R_2 \\
Class\,3 &: R_2 < x_1^2 + x_2^2
\end{aligned}
\tag{40}
$$

Fig. 23 shows the classification of Three-Centered-Ring dataset. This dataset has 300 training samples in each class, and 3000 test samples are considered.

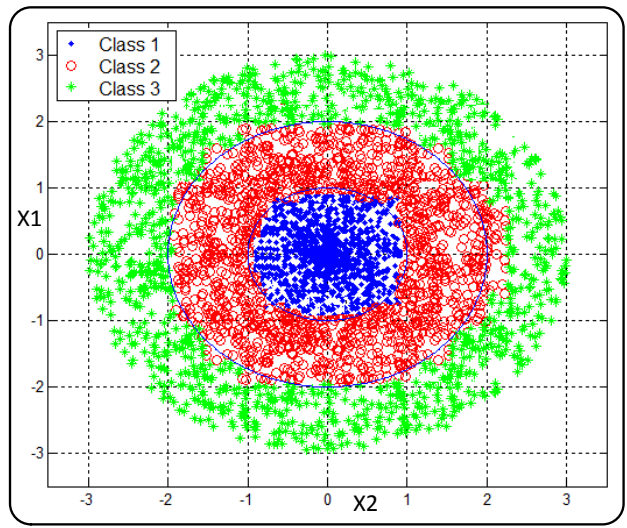

Fig. 23　The classification result on Three-Centered-Ring dataset with $R_1 = 1$ and $R_2 = 4$. Parameters are as follows. $\sigma = 2, \alpha_1 = 0.09, \alpha_2 = 0.27$. 5 partitions are considered on each input domain. The classification accuracy in this experiment is 0.961.

With regard to simulation results, it is apparent that the proposed algorithm employs simpler hardware and software, while performing the learning process with higher speed and comparable precision as ALM. It should also be mentioned that because of small hardware requirements, the proposed algorithm is implementable.

## VII. Conclusion

One of the processing tools in soft computing is ALM-IDS which is inspired by some human brain behavior. Despite its great performance in various applications such as function approximation, classification, etc., it suffers from high computation and hardware complexity that stall its wide popularity. In this paper a novel learning algorithm based on IDS operator was proposed that employs three describing vectors for each IDS plane. The proposed algorithm shows the same performance as ALM , yet with considerable decrease in

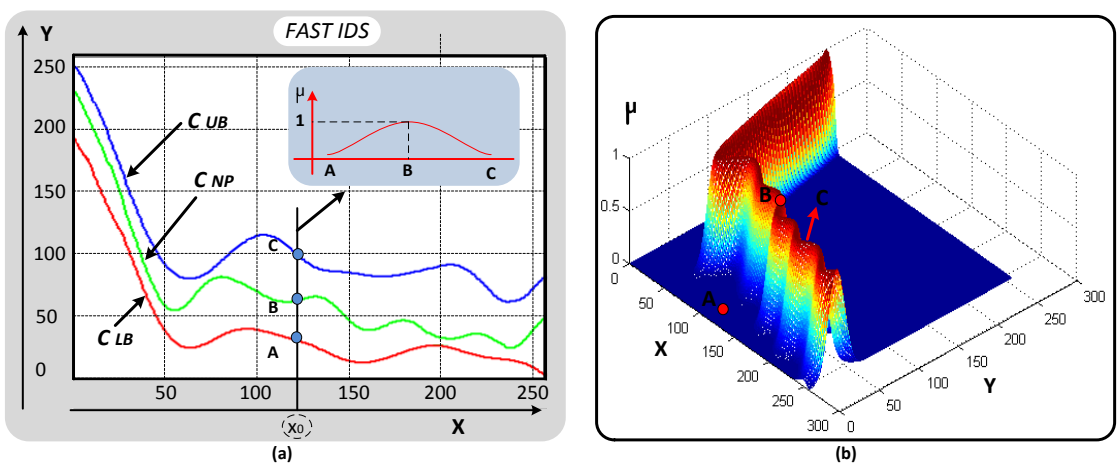

Fig. 24　(a) The fuzzy output of one of the IDS planes associated with $x_i - y$ of function $F_2$ in the proposed algorithm. The associated output of a typical input $x_i$ is fuzzy and any arbitrary membership function can be assigned to it with regard to describing vectors. The center of membership function is located on the *Narrow Path* and it is expanded according to the *Spread* value. In (b) for each input a Gaussian membership function is centered on the *Narrow Path* and its variance is considered to be proportional to the *Spread* value.

process time and hardware size. In this paper, the proposed algorithm was first described in full details and then two applications evaluated its performance. With regard to simulation results it can be mentioned that the proposed algorithm shows great performance and the proposed hardware enjoys smaller and simpler implementation compared to traditional implementations. Therefore, the advantages of the proposed algorithms can be summarized as follows.

1) Appropriate precision and speed.
2) Low implementation cost.
3) Considerable decrease in the number of memristors from $O(n^2)$ to $O(3n)$.
4) Qualitative description of IDS plane is replaced by quantitative one.
5) Despite artificial neural networks whose learned knowledge is not easy to represent, Fast ALM, like ALM, is pattern based and provides useful and easy to understand representation from learned knowledge.
6) Because the CMOS circuit and the memristor crossbar structure are isolated, the proposed hardware is compatible with the memristor / CMOS platform which means it can be implemented with CMOL technology.
7) Nano scale implementation of memristors consumes low power to change their memristance.
8) FPGA or ASIC implementation of ALM are not efficient in terms of hardware size and power consumption.
9) Like ALM algorithm, Fast ALM provides fuzzy output for each IDS plane, but because the upper-bound and the lower-bound are extracted, the fuzzy membership function can be defined on output values as shown in Fig. 24.

Fast ALM satisfies the drawbacks of ALM to a great extent. The proposed algorithm provides a promising tool for various applications such as function approximation, classification, etc. In noisy datasets, multi epoch learning can be employed, which was not available in ALM. By performing multi epoch learning, each sample is introduced to learning algorithm more than once and this reduces the impact of noisy datasets.

Like ALM, Fast ALM also suffers from need to partitioning the inputs domain. Further research in this area is required to propose a novel approach to find the optimal number of partitions and their positions. Optimization algorithms like Genetic Algorithm has been used to perform such partitioning[41].

Another challenge facing researchers in memristive systems is the high computational cost of simulating large memristor crossbar structures. Although the proposed algorithm considerably reduces the required number of memristors, if high resolution process is required, larger memristor crossbar structure requires more powerful processors to simulate such a structure. Because of the high parallelization potential of the proposed algorithm, GPUs and multi-core processors can be utilized to perform parallel processes and reduce the simulation time and computational complexity.

ACKNOWLEDGMENT

All the experiments and ideas of this research work have been developed in Artificial Creatures Lab, belonging to the Electrical Engineering Department, Sharif University of Technology, and Tehran, Iran.

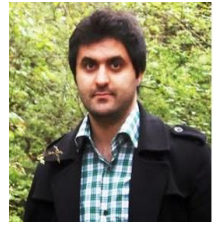

**Sajad Haghzad Klidbary** received the B.Sc. degree in Electrical Engineering in 2009 from Razi University, Kermanshah, Iran, and M.Sc. degree on Digital electronics from Department of Electrical Engineering, Sharif university of Technology, Tehran, Iran in 2012. He is currently PhD candidate of Electrical Engineering in Artificial Creatures Lab, Sharif University of Technology. His research interests include robotics, Artificial Intelligence, Neural Networks, Fuzzy Systems, Genetic and Evolutionary Algorithms, FPGA circuit design, Memristor and Neuromorphic systems implementation based on Memristor.

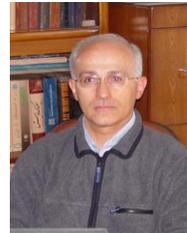

**Saeed Bagheri Shouraki** received his B.Sc. in Electrical Engineering and M.Sc. in Digital Electronics from Sharif University of Technology, Tehran, Iran, in 1985 and 1987. He joined soon to Computer Engineering Department of Sharif University of Technology as a faculty member. He received his Ph.D. on fuzzy control systems from Tsushin Daigaku, Tokyo, Japan, in 2000. He continued his activities in Computer Engineering Department up to 2008. He is currently a Professor in Electrical Engineering Department of Sharif University of Technology. His research interests include control, robotics, artificial life, and soft computing.

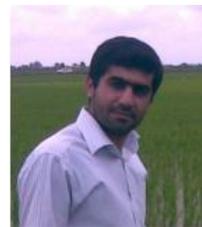

**Iman Esmaili Pain Afrakoti** received his B.Sc. degree in Electrical Engineering in 2006 from Khaje Nasir Toosi University of Technology, Tehran, Iran, and M.Sc and Ph.D degree in Digital Electronic from Department of Electrical Engineering, Sharif University of Technology, Iran in 2008 and 2014 respectively. He is currently an assistant Professor in Engineering and Technology Department of University of Mazandaran. His research interests include Memristor, Neuromorphic systems implementation based on Memristor crossbar structures, ASIC and FPGA, Fuzzy Systems, Artificial Neural Networks and Pattern Recognition.